\documentclass[a4paper,10pt,twoside]{cpc-hepnp}

\usepackage{multicol}
\usepackage{graphicx}
\usepackage{booktabs}
\usepackage{amssymb,bm,mathrsfs,bbm,amscd}
\usepackage[tbtags]{amsmath}
\usepackage{lastpage}
\usepackage{epstopdf}

\begin{document}

\fancyhead[c]{\ Submitted to Chinese Physics C}

\fancyfoot[C]{\small 010201-\thepage}

\footnotetext[0]{Received 6 December 2015}

\title{Discriminating cosmic muon and x-ray based on rising time using GEM detector\thanks{Supported by National Natural Science
Foundation of China (Grants No. 11135002, No. 11275235, No. 11405077 and No.11575073) }}
\author{ WU Hui-Yin$^{1}$
\quad ZHAO Sheng-Ying$^{1}$
\quad WANG Xiao-Dong$^{2}$
\quad ZHANG Xian-Ming$^{1}$ \\
\quad QI Hui-Rong$^{3}$
\quad ZHANG Wei$^{1}$
\quad WU Ke-Yan$^{1}$\\
\quad HU Bi-Tao$^{1}$
\quad ZHANG Yi $^{1;1}$\email{yizhang@lzu.edu.cn}
}

\maketitle

\address{%
$^1$ School of Nuclear Science and Technology, Lanzhou University, Lanzhou 730000, China\\
$^2$ School of Nuclear Science and Technology, University of South China, Hengyang 421001, China\\
$^3$ State Key Laboratory of Particle Detection and Electronics, Beijing 100049, China

}

\begin{abstract}
Gas electron multiplier(GEM) detector is  used in Cosmic Muon Scattering Tomography and neutron imaging in the last decade. In this work, a triple GEM device with an effective readout area of 10 cm $\times$ 10 cm is developed, and an experiment of discriminating between cosmic muon and x-ray based on rising time is tested. The energy resolution of GEM detector is tested by $^{55}$Fe ray source to prove the GEM detector has a good performance. The analysis of the complete signal-cycles allows to get the rising time and pulse heights. The experiment result indicates that cosmic muon and x-ray can be discriminated with an appropriate rising time threshold.
\end{abstract}

\begin{keyword}
Gas Electron Multiplier, Cosmic Muon, X-ray, Rising Time, Discrimination
\end{keyword}

\begin{pacs}
29.40.Gx, 29.40.Cs, 07.05.Tp
\end{pacs}

\footnotetext[0]{\hspace*{-3mm}\raisebox{0.3ex}{$\scriptstyle\copyright$}2013
Chinese Physical Society and the Institute of High Energy Physics
of the Chinese Academy of Sciences and the Institute
of Modern Physics of the Chinese Academy of Sciences and IOP Publishing Ltd}%

\begin{multicols}{2}

\section{Introduction}

The intention of the GEM detector is to detect charged particles with good performance, such as high effective gain, good position resolution, high counting rate and so on~\citep{lab1,lab2}. GEM detector can also detect ray with a gain as high as 10$^6$ for triple GEM foils. Due to the good position resolution and high efficiency of detection, GEM detector is widely used in Cosmic Muon Scattering Tomography(MT)~\citep{lab3}. In most cases, the bottleneck for cosmic tomography is the flux. Usually the flux of cosmic muon  is about 1 cm$^{-2}$ minute$^{-1}$, so the exposure time is crucial in practical application. Background induces a great impact on the minimum exposure time and track reconstruction, therefore reducing the background is particularly important. Taking into account the tomography environment and GEM detection efficiency of different rays, the largest contribution of the background is x-ray. Furthermore GEM detector is also applied on neutron imaging extensively with its plasticity. GEM detector can easily become a neutron detector by coupling with High-Density Polyethylene (HDPE) as a neutron-proton converter~\citep{lab4,lab5}. The x-ray generated by neutron have a great interference the measurement of the effective neutron response~\citep{lab6} and proton track reconstruction. Neutron response signals and x-rays are discriminated by the pulse heights in Ref.~\citep{lab6}, with a big deviation.
As the same time, GEM detector is widely used in high energy physics experiment as a track detector. The x-ray have an influence on the track reconstruction of high energy particles which will run through the GEM detector.
\\
\indent
  Therefore, discriminating x-ray and cosmic muon is meaningful for muon tomography and neutron imaging. The traditional method of reducing the x-ray is anti-coincidence with two plastic scintillators, which complicates the experimental setup. Discriminating x-ray and charged particle is also a meaningful work.

\indent
  In gas, different rays deposit energies in different ways. For charged particle, such as proton, it deposits energy along its track and produces ion-electron pairs over the whole track. As a contrast, x-ray loses its energy in a small region, different ways of depositing energy present differently on the time characters of signals. In this paper, we discriminate cosmic muon which deposits energy along its track like proton and x-ray based on the rising time. The conclusion can also be generalized to discriminating between charged particle and photon in MPGD.

\section{Experimental setup for discriminating between cosmic ray and x-ray}

The triple GEM detector works in a proportional mode as shown in Fig.~\ref{fig1}. It consists of three GEM foils, a cathode plane and a read-out anode with 496 one-dimensional strips. The sensitive area is 10 cm$\times$10 cm and the width of the strips is 100 $\mu$m. Three GEM foils separate the chamber into four gaps. The thickness of the drift gap is 3 mm, while the thickness of induction gap is 4 mm. The distance between the two GEM foils is 2 mm. The detector is operated based on a continuously flushed Ar/CO$_2$ gas mixture (80/20 percentage in volume). A voltage divider is employed to supply the bias voltage to the detector, which avoid electric fields becoming too high in the case of a discharge. For the further protection, an additional 10 M$\Omega$ protection resistor is connected with the electrode. The work bias voltage is -2825 V, while the bias voltage of GEM foil is 353 V, and the electric field of drift gap and induction gap are 1177 V/cm and 1324 V/cm respectively.
\begin{center}
\includegraphics[width=8cm]{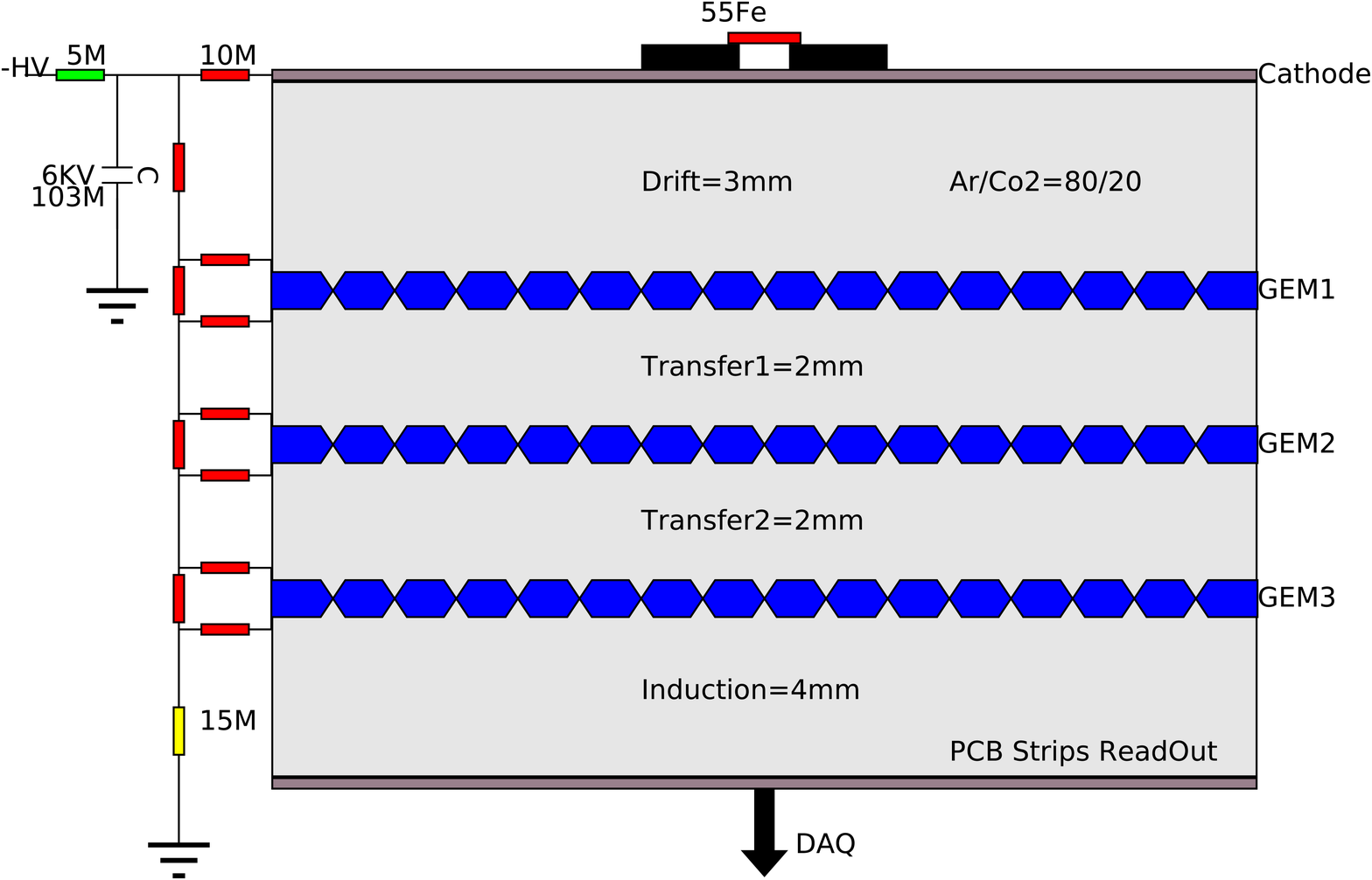}
\figcaption{\label{fig1}    Experimental principal scheme of triple GEM detector. }
\end{center}
\par
\indent
The energy spectrum of the triple GEM detector is measured with $^{55}$Fe 5.9keV x-ray source shown in Fig.~\ref{fig2}. The energy resolution is about 20.4\%, the escape peak of Argon atom and full photo-electron peak of $^{55}$Fe x-ray is distinguished completely, which demonstrates the GEM detector has a good performance. The detection system has a good energy linearity relationship, for the ratio of two peaks is 1.96.

\par
\indent
Experimental setup consists of two plastic scintillators with 10 cm$\times$15 cm effective area and a GEM detector placed between two scintillators. In order to conduct signal studies, The plastic scintillators signals and the GEM detector signals pre-amplified by a charge sensitive preamplifier (ORTEC 142B) were recorded, using a 1.8 GHz sample frequency FlashADC with four channels. The working voltage of the scintillator is set to a suitable value so that the height of signal of cosmic muon can be recorded and easy to compare with a threshold. The work bias voltages of plastic scintillators are -1483 V and -1273 V, respectively. The bias voltages are delivered to each detectors by CAEN module N472.

\begin{center}
\includegraphics[width=8cm]{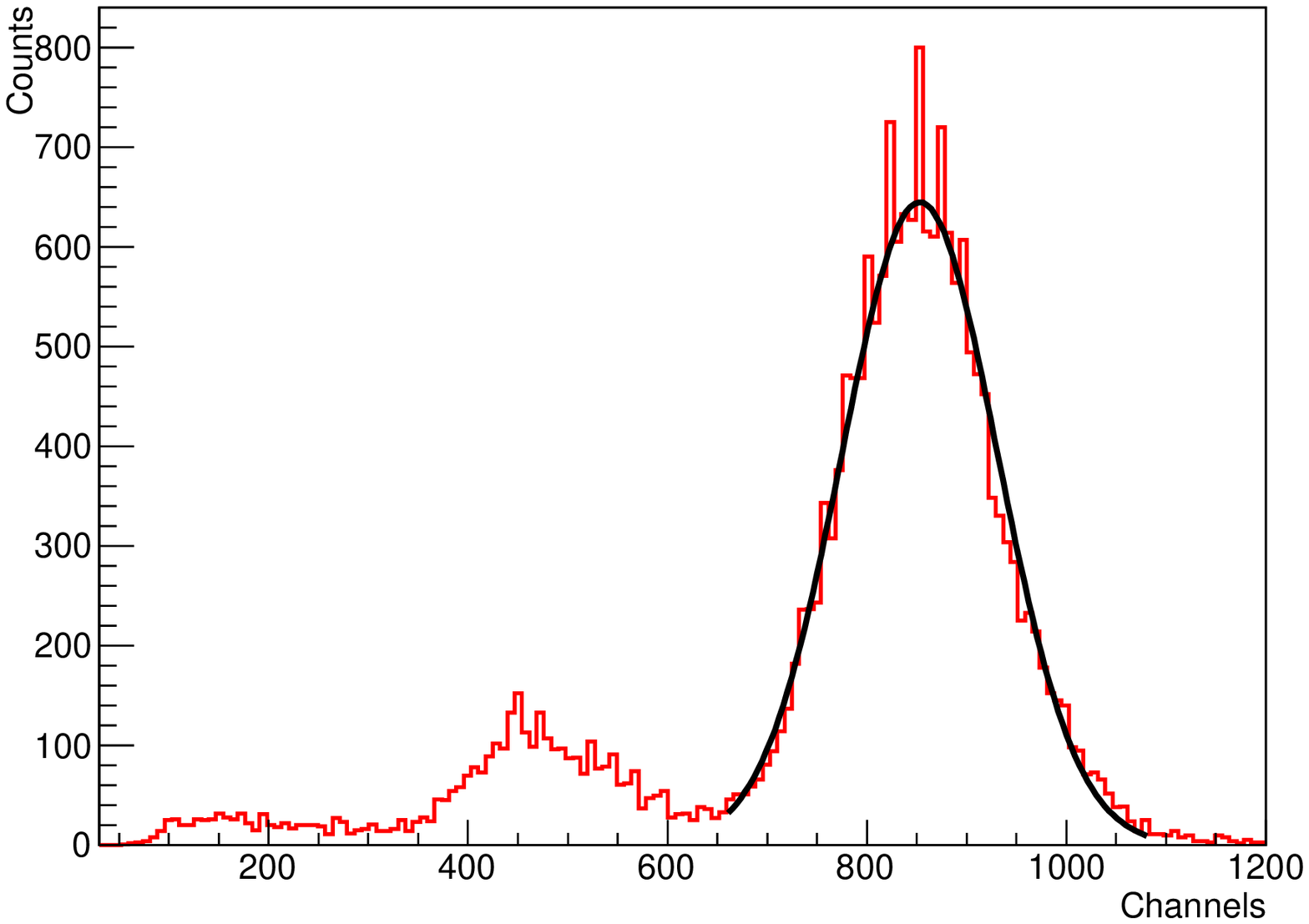}
\figcaption{\label{fig2}    $^{55}$Fe 5.9 keV x-ray source spectrum measured by triple GEM detector. }
\end{center}
\begin{center}
\includegraphics[width=8cm]{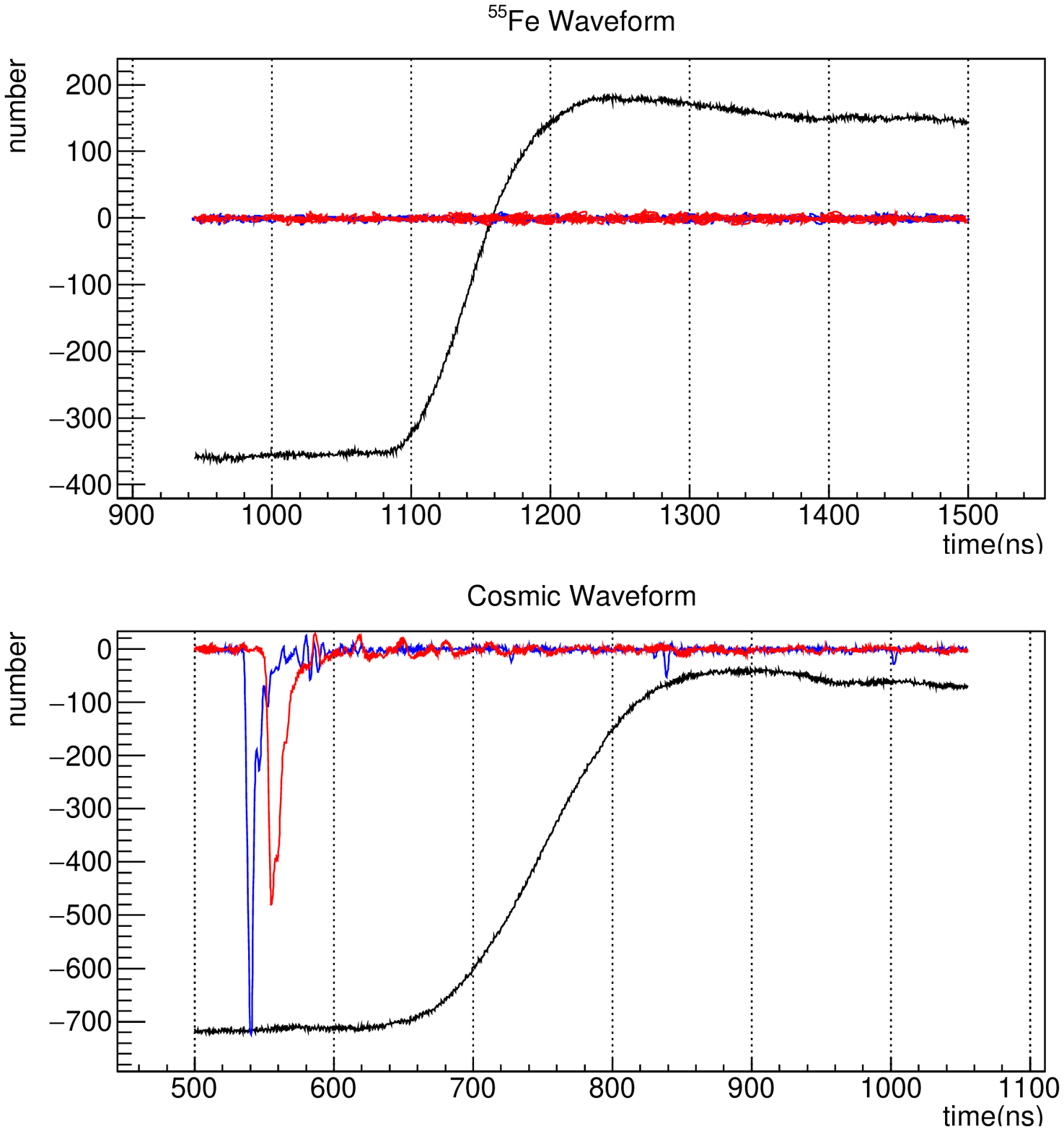}
\figcaption{\label{fig3}  (color online)  Typical signals of three detectors. The blue curve sample and red curve sample represent the signals of top plastic scintillator and bottom plastic scintillator, respectively. The black curve sample is the signal of GEM detector from x-ray or cosmic muon. }
\end{center}
\par
\indent
Two typical signals are shown in Fig.~\ref{fig3}, black curve is the GEM detector signal waveform of $^{55}$Fe x-ray or cosmic muon, blue curve sample represents the ray signal waveform of top plastic scintillator, and the red curve sample represents the ray signal waveform of bottom plastic scintillator. $^{55}$Fe x-ray signals were triggered by internal trigger using GEM detector signal, while the cosmic muon signals were triggered by internal trigger using signals from the plastic scintillator. The signals are easily discriminated by analyzing signals from the three detectors with coincidence and anti-coincidence method. If the magnitudes of two plastic scintillators are both larger than a given threshold in one event, the corresponding signal of GEM detector is identified as a cosmic muon signal. In contrast, the signal of GEM is identified as induced by x-ray if the magnitudes of two plastic are both smaller than a given threshold.

\section{Results and Discussion}
Fig.~\ref{fig4} shows the energy deposition of cosmic muon (blue curve) and the energy spectrum of $^{55}$Fe 5.9 keV x-ray (red curve), cosmic muon spectrum fits well to a typical Landau distribution (black curve) as expected from minimum ionizing particles, it imply that the events discriminated by coincidence are indeed the cosmic muon events. With the definitions that the detector's rising time is defined by the time span required for the output signal to rise from 10\% to 90\% level of the maximum value, the result of the rising time for $^{55}$Fe x-ray and cosmic muon with exponent coordinate of Y axis is shown in Fig.~\ref{fig5}. The total counts of cosmic muons equal to the total counts of $^{55}$Fe x-rays. It has been determined by fitting gaussian function in the $^{55}$Fe x-ray rising time spectrum and cosmic muon rising time spectrum. The FWHM of rising time for 5.9 keV x-ray is 4.11 ns and for cosmic muon is 53.69 ns, the mean of rising time for 5.9 keV x-ray is 89.4 ns, while the mean of rising time for the cosmic muon is 135.7 ns.
\begin{center}
\includegraphics[width=8cm]{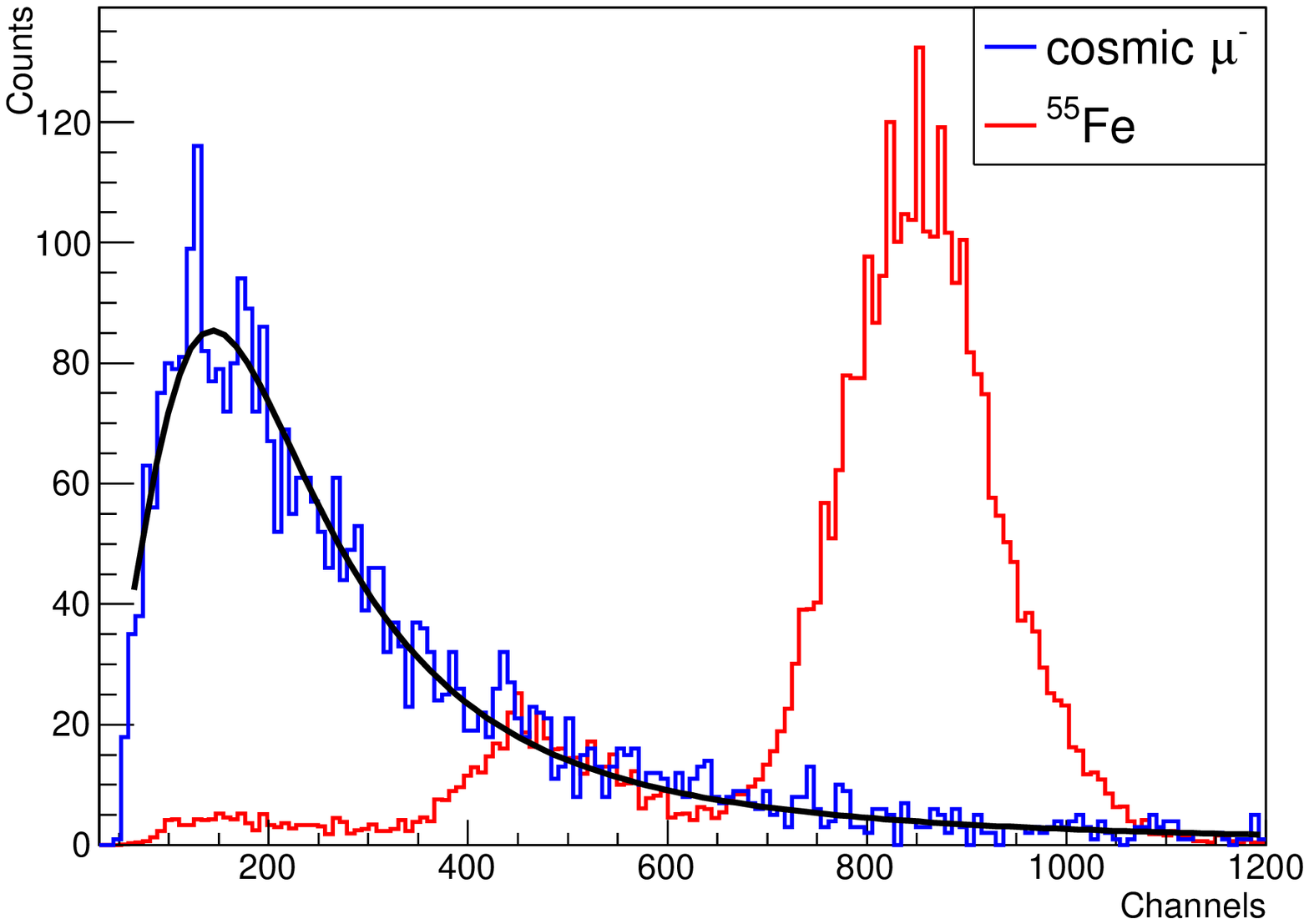}
\figcaption{\label{fig4}   (color online) Energy deposition of cosmic muon (Blue curve) and  energy spectrum of $^{55}$Fe x-ray (Red curve). The black line sample represent the typical Landau distribution fitted by cosmic muon spectrum. }
\end{center}

\par
\indent
There are two reasons to account for rising time of 5.9 keV x-ray. First, electron drift time in the induction gap (4 mm) is main contribution for the rising time of 5.9 keV x-ray. Fig.~\ref{fig6} shows a simulation of charge induction in the readout PCB with 4 mm induction gap based on GARFIELD++~\citep{lab7}. The induction gap contributes about 74 ns, according to our simulation. Second, the rising time of the preamplifier (ORTEC 142B) have a great influence upon the rising time of signal. However, with the increase of the detector output capacitance, the rising time increases.
\begin{center}
\includegraphics[width=9cm]{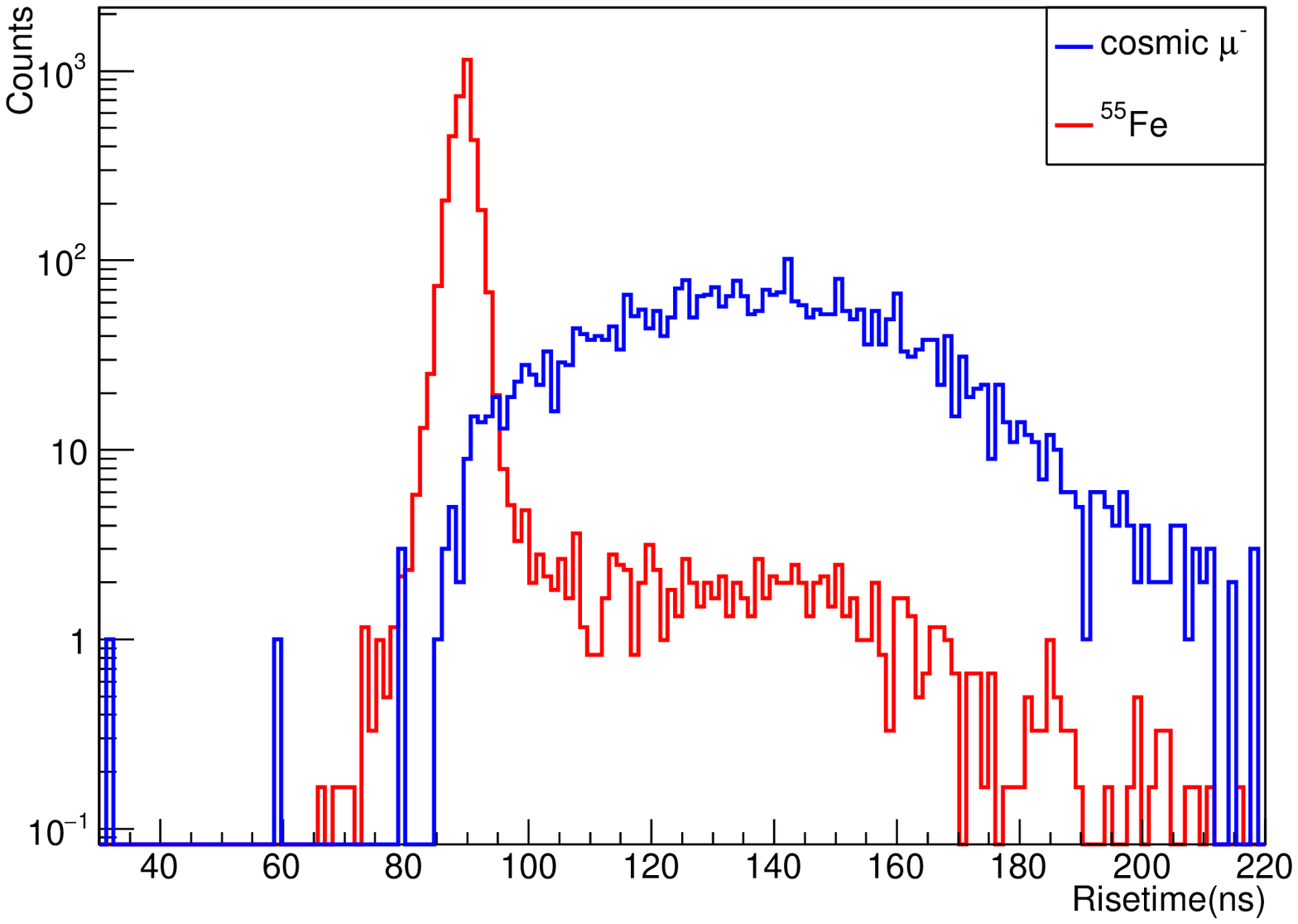}
\figcaption{\label{fig5}   (color online) The rising time spectrum for $^{55}$Fe 5.9 keV x-ray and cosmic muon. The blue line sample and red line sample represent the rising time of cosmic muon and the rising time of x-ray respectively.}
\end{center}

\begin{center}
\includegraphics[width=8cm]{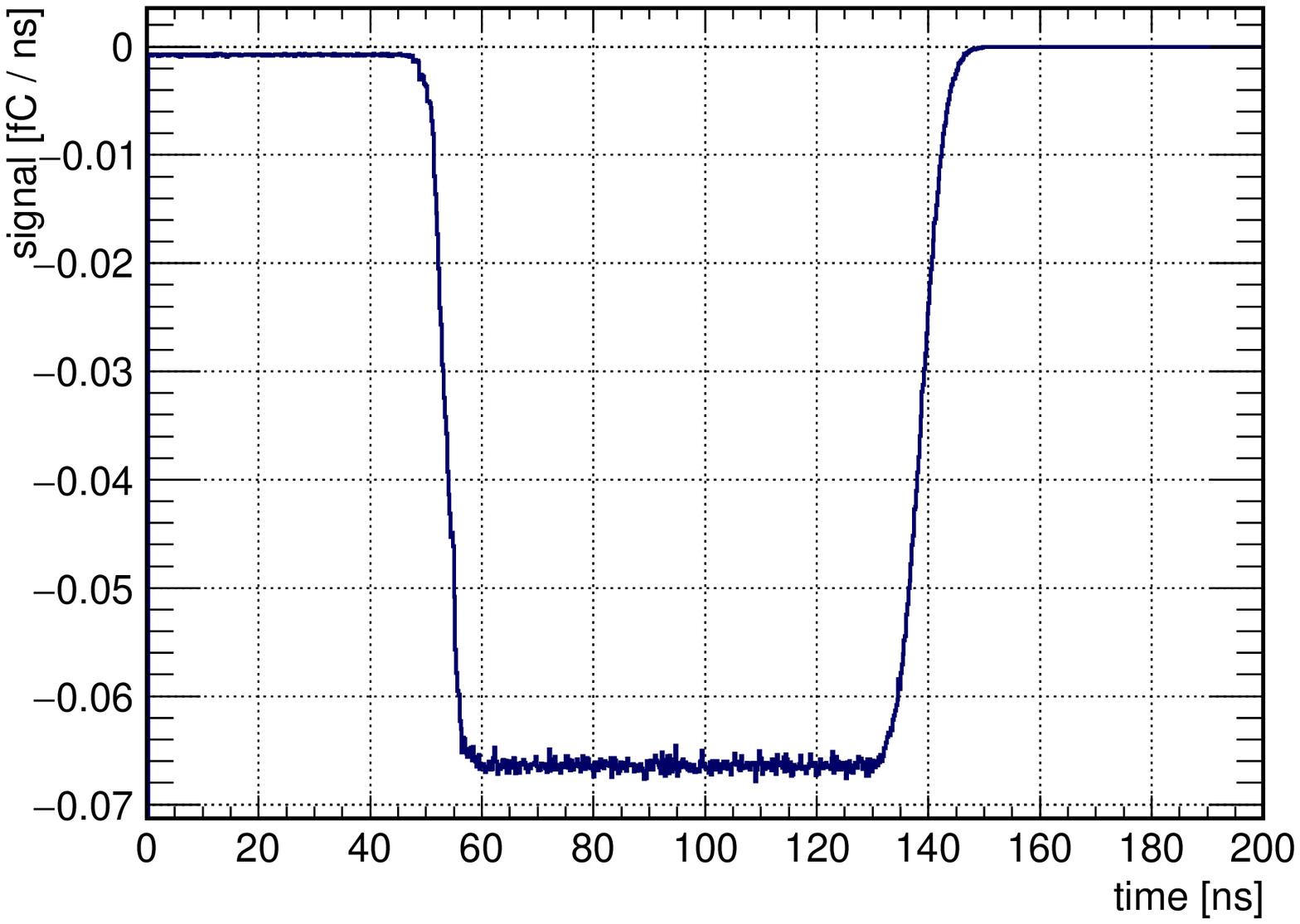}
\figcaption{\label{fig6}    Result of the simulation of charge induction in the readout PCB with 4mm induction gap based on GARFIELD++.}
\end{center}
\par
\indent
The FWHM of x-ray is smaller than the FWHM of cosmic muon because the ion-electron pairs produced by x-ray is in a small region, its electron density is larger than the density of electron produced by cosmic muon which deposit its energy over its track. All electrons from x-ray drift into induction gap in a small time scale, with large electron density, which can immediately respond the preamplifier. Electrons from cosmic muon drift into induction gap in a lager time scale based on the thickness of drift gap, with small electron density, which lead to the discreteness of the preamplifier response time. The large FWHM of cosmic muon's rising time decrease effect on discriminating x-ray and cosmic muon, especially on the conditionb that the rising time of x-ray is close to the rising time of cosmic muon.

\par
\indent
There are some x-ray events with rising time longer than 100 ns in the $^{55}$Fe x-ray rising time spectrum shown in Fig.~\ref{fig5}. These events do not meet the statistical law (Gaussian statistics) obviously. Taking into account the experimental set-up, the effective area of both plastic scintillators are 10 cm$\times$15 cm, closing to the effective area of the
GEM detector, so the scintillator can not completely cover the GEM detector. These events are suspect to be the cosmic muons which do not through the top scintillator
or bottom scintillator. In Fig.~\ref{fig7}, the samples (red solid circle) represent the percentage of the total counts of $^{55}$Fe x-ray events, whose rising time are longer than the threshold, and the samples (blue solid triangle) represent the percentage of the total counts of cosmic muon events, whose rising time are shorter than the threshold. 97\% of cosmic muons can be discriminated from x-rays, and 97\% of x-rays can be discriminated from cosmic muons when the threshold is 97 ns.
\begin{center}
\includegraphics[width=8cm]{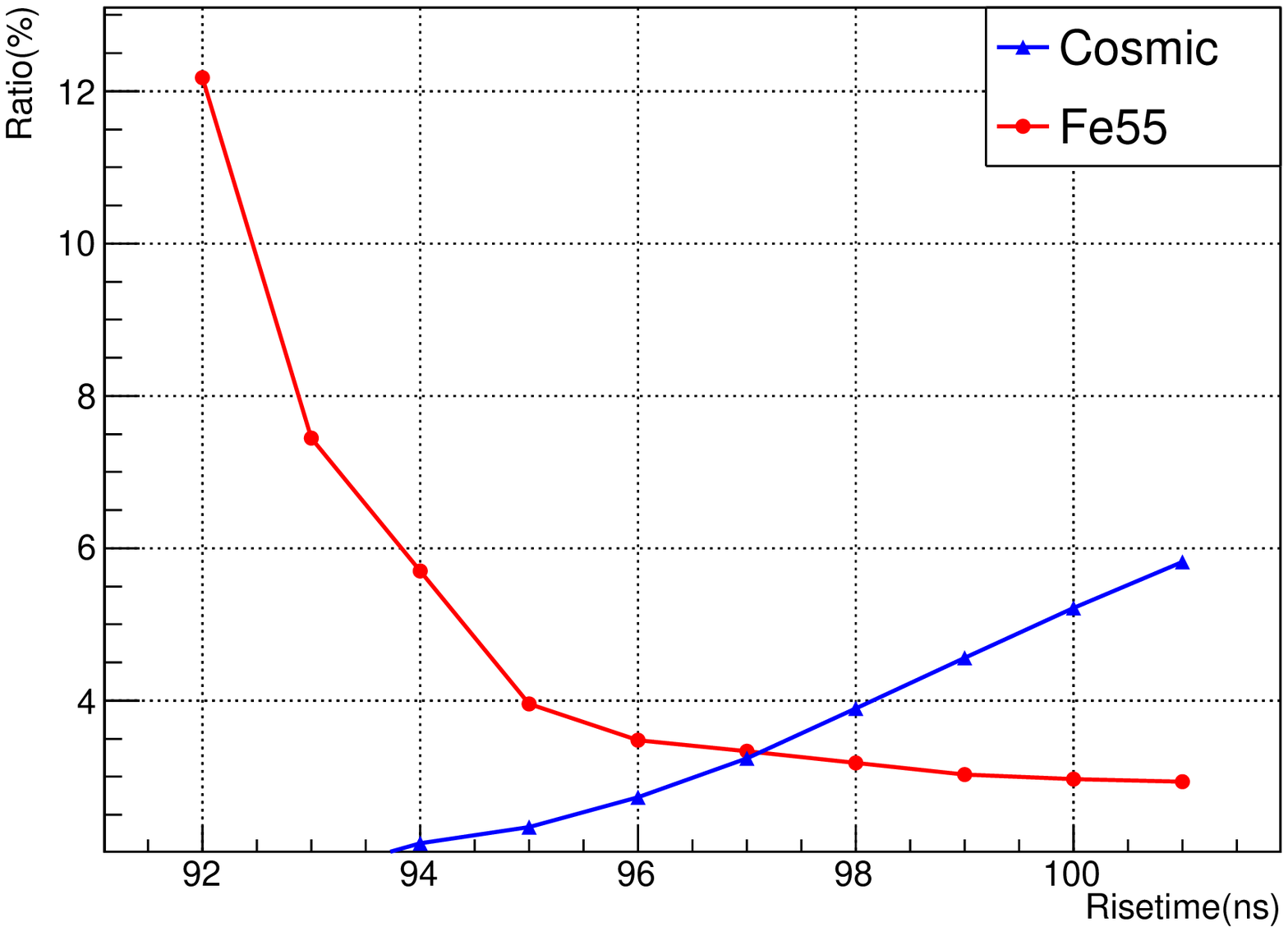}
\figcaption{\label{fig7}  (color online)  Discriminated result of $^{55}$Fe x-ray(red solid circle) and cosmic muon(blue solid triangle) with different threshold. }
\end{center}
\section{Conclusions}
Energy resolution of the triple GEM detector is measured by $^{55}$Fe 5.9 keV x-ray source, the results confirm that the triple GEM detector has a good energy resolution around 20.6\% and
a good energy linearity relationship. The discriminating experiment based on rising time was studied, experiment results confirm that 97\% of cosmic muon and x-ray can be discriminated through rising time for GEM detector. We can infer that the rising time of signal can also be used to discriminate charged particles and photons. This method can improve neutron detection data and improve the track accuracy of charged particle in high energy physics experiment. For the further optimization, three measures can be taken to reduce the intrinsic rising time. First, using fast time preamplifier instead of 142B, such as APV25~\citep{lab8}. Second, As mentioned before, reducing the capacitance can reduce the rising time. So interconnecting less readout strips and grounding the other strips can reduce the capacitance of GEM detector with the sacrifice of effective area. Finally, reducing the thickness of induction region with a 100 $\mu$m-thick metallic mesh. However, it is easy to increase cosmic muon rising time by increasing the thickness of drift region and decrease the FHMW of cosmic muon by increasing the voltage on GEM foil.
\par
\indent
As discussed above there are two important things should to do in the next work. They are that the construction of the GEM detector should be improved with the methods mentioned before to discriminate x-ray and cosmic muon better, and neutron source should be detected to discriminate x-ray and recoil proton based on rising time.\vspace{4ex}


\acknowledgments{ We are grateful to researcher DUAN Li-Min, associate researcher Hu Rong-Jiang, Yang He-Run, Dr.LU Chen-Gui and Dr.Zhang Jun-Wei for
their helpful discussions and support in Institute of Modern Physics, Chinese Academy of Sciences.}

\end{multicols}


\vspace{-1mm}
\centerline{\rule{80mm}{0.1pt}}
\vspace{2mm}



\clearpage

\end{document}